\documentclass[submission, Phys]{SciPost}

\usepackage{subcaption}
\usepackage{fancyhdr}
\begin{document}
\fancyhf{}
\fancyhead{} 
\renewcommand{\headrulewidth}{0pt}
\begin{center}{\Large \textbf{Consistency Relation for Fixed Point Dynamics} 
}\end{center}

\begin{center}
Oleg Antipin\textsuperscript{1},
Alan Pinoy\textsuperscript{2},
Francesco Sannino\textsuperscript{3,4,5},
Shahram Vatani \textsuperscript{6*}
\end{center}

\begin{center}
{\bf 1} Rudjer Boskovic Institute,
  Division of Theoretical Physics,
  Bijeni\v cka 54, 10000 Zagreb, Croatia
\\
{\bf 2} Dept. de Mathématique, Université Libre de Bruxelles, Boulevard du Triomphe, Bruxelles, Belgique
\\
{\bf 3} Quantum  Theory Center ($\hbar$QTC) \& D-IAS, Southern Denmark Univ., Campusvej 55, 5230 Odense M, Denmark\\ 
{\bf 4}   Dept. of Physics E. Pancini, Università di Napoli Federico II, via Cintia, 80126 Napoli, Italy
\\
{\bf 5} INFN sezione di Napoli, via Cintia, 80126 Napoli, Italy
\\
{\bf 6} Centre for Cosmology, Particle Physics and Phenomenology (CP3), Université catholique de Louvain, Chemin du Cyclotron, 2 B-1348 Louvain-la-Neuve, Belgium

\end{center}

\centerline{\bf Abstract}
\vskip .3cm
{
 We gain insight on the fixed point dynamics of  $d$ dimensional quantum field theories by exploiting the critical behavior of the $d-\epsilon$  sister theories. To this end we first derive a self-consistent relation between the $d-\epsilon$ scaling exponents and the associated $d$ dimensional beta functions. We then demonstrate that to account for an interacting fixed point in the original theory the related $d-\epsilon$     scaling exponent must be multi-valued  in  $\epsilon$.  We elucidate our findings by discussing several examples such as the QCD Banks-Zaks infrared fixed point, QCD at large number of flavors, as well as the O(N) model in four dimensions. For the latter, we show that although the $1/N$ corrections prevent the reconstruction of the renormalization group flow, this is possible when adding the  $1/N^2$ contributions.
}
\newpage

\section{Introduction}

Solving for the gauge dynamics of quantum field theories (QFT)s is paramount to properly describe the dynamics of different critical phenomena from cosmology to particle and condensed matter physics. Of particular interest are quantum field theories developing fixed points (FP)s dynamics. Here the theories are typically conformal and display universal behaviors captured by their conformal data such as scaling exponents. 

It is therefore interesting to explore relations among different QFTs. These can emerge in several ways, from gauge-gauge dualities \cite{Seiberg:1994aj,Seiberg:1994bz,Seiberg:1994pq,Seiberg:1994rs,Intriligator:1995au} to holography \cite{Maldacena:1997re,Witten:1998qj}. Another way, the one we investigate here, is to relate QFTs across different space-time dimensions \cite{Gracey:1993sn,Gracey:1993kc,Gracey:1990wi,Gracey:2018qba,Gracey:2018ame}. Here the idea is to derive exact non-perturbative relations between scaling exponents in $d-\epsilon$ dimensional theories and  the  $d$ dimensional beta functions of associated QFTs. The resulting self-consistent relation will allow us to show that, in order to capture the infrared dynamics of an interacting fixed point for the $d$-dimensional theory, the scaling exponent of the $d-\epsilon$ sister theory must be multi-valued in $\epsilon$. More generally we will show that is possible to capture the overall behavior of the $d$-dimensional beta function in terms of the multi-valued nature of the scaling exponent of the $d-\epsilon$ sister theory. 

Various time-honored examples are considered to highlight different features of the self-consistent relation. For example, the Banks-Zaks \cite{Banks:1981nn}  perturbative nature of the Quantum Chromodynamics (QCD)  infrared fixed point is reflected in the double valued behavior of the $4-\epsilon$ critical exponent. We further show that for QCD at large number of flavors the single valued nature of the $d-\epsilon$ scaling exponent prevents to determine whether an ultraviolet fixed point is excluded, contrary to the statement made in \cite{Alanne:2019vuk} but in agreement with \cite{Sannino:2019vuf}. For the $O(N)$ model we demonstrate that although the $1/N$ corrections prevent the reconstruction of the renormalization group (RG) flow, this is possible when adding the  $1/N^2$ contributions.

The paper is organized as follows. In Sec.\ref{SCRcondition} we derive the self-consistency relation between the critical exponent and the beta function of the theory and study its general properties. In Sec.\ref{applications} we apply our findings to the gauge theories at one- and two-loop order and to the gauge and $O(N)$ models with
a large number of degrees of freedom $N$, in which
perturbative expansion for the RG functions is re-organized in powers
of $1/N$. We present our conclusions in Sec.\ref{conclusions}.

 \section{The Self Consistency Relation } \label{SCRcondition}
We consider a single coupling QFT in dimensions $d-\epsilon$ where $d$ is the critical dimension of its bare coupling $\alpha_0$ with the associated operator being marginal. In the process of renormalization one introduces RG scale $\mu$ and $Z$-factors relating the
renormalized fields and coupling $\alpha$ to the bare ones. Imposing the bare coupling to be $\mu$-independent, 

\begin{equation*}
    \mu \frac{ \text{d}}{\text{d}\mu} \left[ \alpha \; Z \; \mu^{\epsilon} \right]=0
\end{equation*}
yields the beta function:

\begin{equation}    \Delta\left(\alpha,\epsilon\right)= \frac{\mathrm{d}\; \alpha}{\mathrm{d} \ln{\mu}}= -\alpha \epsilon + \beta\left(\alpha\right) \label{Beta}
\end{equation}
where $\beta\left(\alpha\right)$ denotes the beta function in $d$ dimensions.

 At criticality the deformed theory reaches a fixed point $\alpha^*$, i.e. $ \Delta\left(\alpha^*, \epsilon \right)=0 $ with the critical exponent $\omega$ defined by the slope of the beta function:
\begin{equation}
\omega\left(\epsilon\right)   \equiv \frac{\partial \Delta\left(\alpha, \epsilon \right)}{\partial \alpha}\Bigr|_{\substack{\alpha^*}}=-\epsilon+\beta'(\alpha^*)  \ . \label{omega}
\end{equation}
Combining Eq.\eqref{Beta} and Eq.\eqref{omega} we arrive at\footnote{From here on we drop the superscript $^*$ on $\alpha$ as we always work at criticality.}: 
\[
\omega\!\Bigl(\tfrac{\beta(\alpha)}{\alpha}\Bigr) 
\;=\;
-\,\tfrac{\beta(\alpha)}{\alpha} + \beta'(\alpha)\,
\]
so that applying the chain rule leads to
a first-order nonlinear differential equation, which we refer to as the \emph{Self-Consistency Relation} (SCR):
\begin{equation}
\label{MasterEqSimp}
\frac{dY(\alpha)}{d\alpha}
\;=\;
\frac{\omega\bigl[Y(\alpha)\bigr]}{\alpha} \,
\end{equation}
where we defined \(Y(\alpha) \equiv \beta(\alpha)/\alpha\).

\subsection{General Features}

We point out that SCR is \emph{non-perturbative}, as no assumptions on the specific form of \(\beta(\alpha)\) have been made beyond the single-coupling renormalizability, and \emph{scheme-independent} since critical exponents are universal physically measurable quantities.

In order to capture the overall behavior of its solutions we first recall two important points about the class of differential equations to which SCR belongs:
\begin{itemize}
    \item Any solution is characterized by an initial condition.
    \item All the solutions $Y$ are monotonic; moreover, depending on the sign of $\omega$, they are either attracted or repelled by the closest zero of $\omega$.
\end{itemize}

\begin{figure}[h!]
    \centering
    \includegraphics[trim={1cm 0cm 0 0},width=0.4\linewidth]{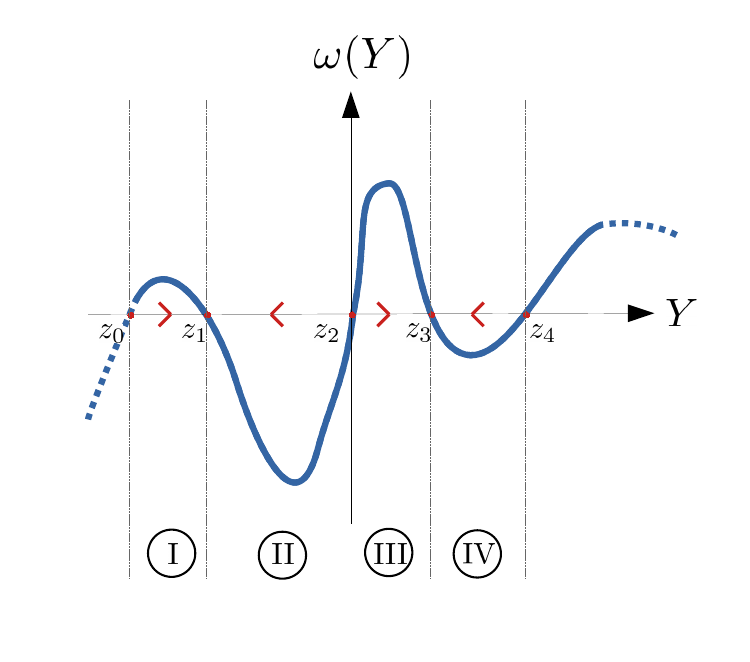} \hspace{1cm}
    \includegraphics[width=0.5 \linewidth]{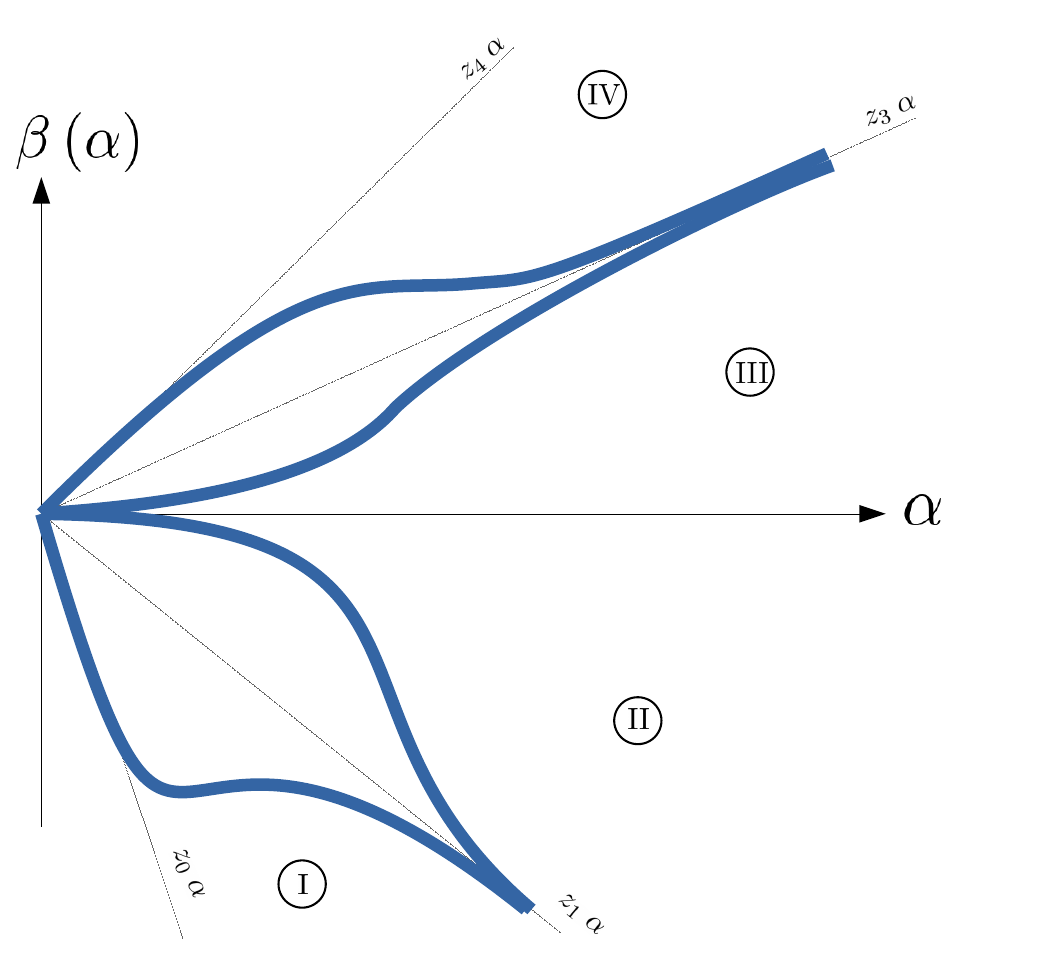}
    \vspace{0.5cm}
    \caption{ (Left) Sketch of $\omega$ with zeroes $z_i$  delimiting the regions \MakeUppercase{\romannumeral 1}, \MakeUppercase{\romannumeral 2}, \MakeUppercase{\romannumeral 3} and \MakeUppercase{\romannumeral 4}. Red arrows indicate the direction of the RG group flows from UV to IR.  (Right) The form of the solutions taken with initial conditions in each region.} 
    \label{phasespace}
\end{figure}

We display these features in Fig. \ref{phasespace} demonstrating that one can sketch the asymptotics of the solutions just by listing the zeros $z_i$ of $\omega$ and specifying an initial condition.  We are assuming $\omega$ is \emph{Lipschitz continuous} to make sure that at $\alpha=0$ and $\alpha=\infty$ the solutions $Y$ flow \emph{asymptotically} to $z_i$. Thus the resulting $\beta$-function behave asymptotically as a straight line : $\lim\limits_{\alpha\to 0}\,\beta(\alpha) \;=\; z_i\,\alpha $  and $ \lim\limits_{\alpha\to \infty}\,\beta(\alpha) \;=\; z_j\,\alpha $ with the only exceptions for the solutions in regions \MakeUppercase{\romannumeral 2} and \MakeUppercase{\romannumeral 3} which behave quadratically \(\beta(\alpha)\sim \beta_0\,\alpha^2\) close to $\alpha=0$ since the linear term vanishes ($z_2=0$).

\vspace{0.25cm}

This quadratic behavior is familiar from the perturbative expansion in QFT around \(\alpha=0\) with \(\beta(\alpha)\sim \beta_0\,\alpha^2\)  which implies $Y(0)=0$.  Combined with the monotonicity of $Y$ it means that \(Y(\alpha)\) cannot vanish at nonzero \(\alpha\) and thus additional \emph{interacting} fixed point(s) in $d$-dimensions cannot exist which is manifest in Fig.\ref{phasespace}.

Demanding such an interacting FP(s) to exist requires to relax the implicit assumption that $\omega(0)$ is a single-valued function. In fact, if equation $\beta(\alpha_i)=0$ has $i$ non-trivial solution(s) means that the criticality condition $\epsilon=\,\tfrac{\beta(\alpha_i)}{\alpha_i}$, in addition to the trivial Gaussian FP solution $(\epsilon,\alpha)=(0,0)$, has another solution(s) $(0,\alpha_i)$ and since $\omega$ for these solutions is, in general, different means that $\omega(0)$ is multi-valued. To illustrate the multi-valuedness of $\omega(\epsilon)$  in $d-\epsilon$ dimensions, in the next subsection, we exploit  the graphical approach where we study the branch of the beta function connected to the Gaussian FP.

\subsection{Non-Lipschitz Behavior and Interactive Fixed Point}
\label{Geometry}

\begin{figure}[h!]
    \includegraphics[width=.4\textwidth]{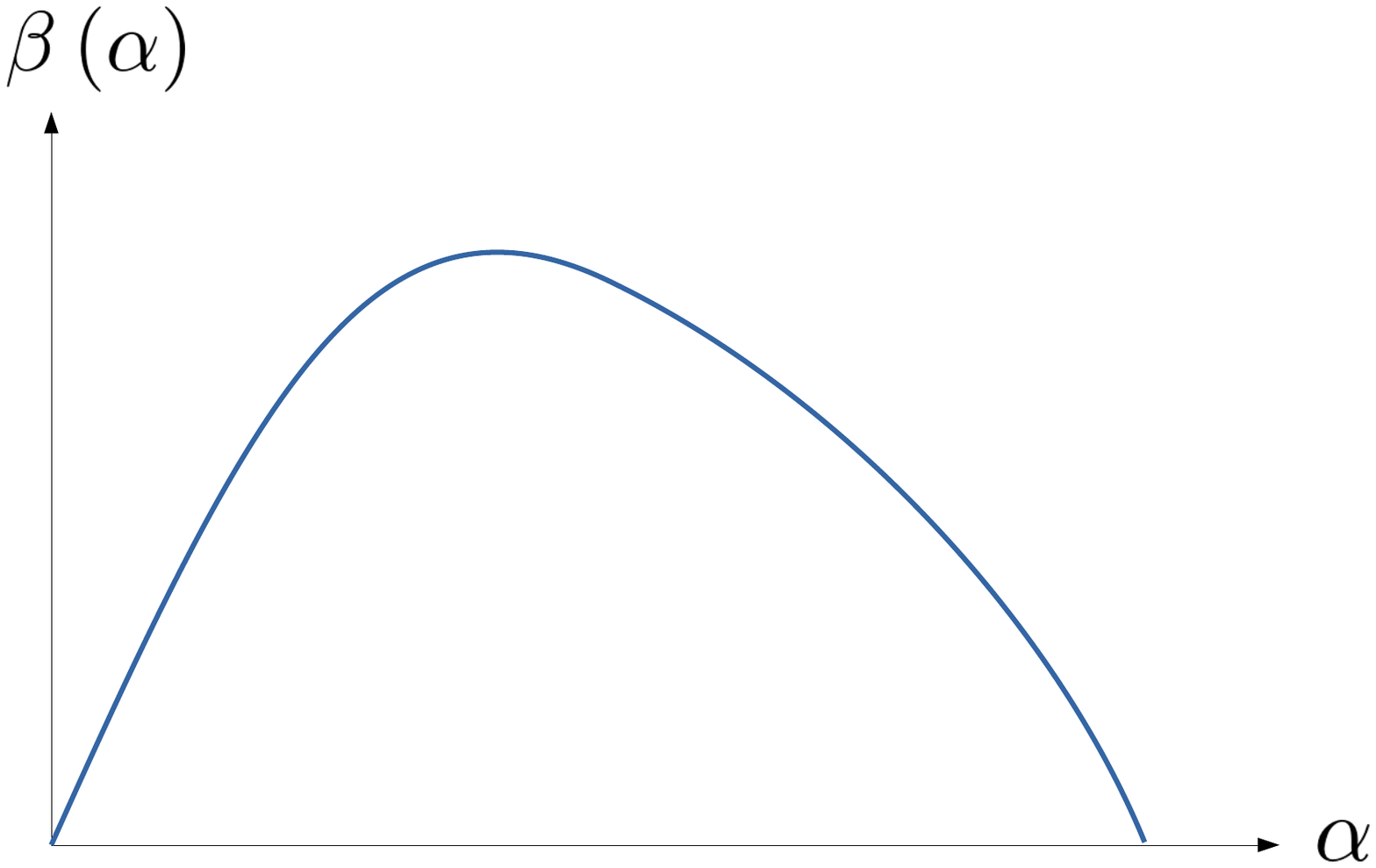}
    \includegraphics[width=0.5\textwidth]{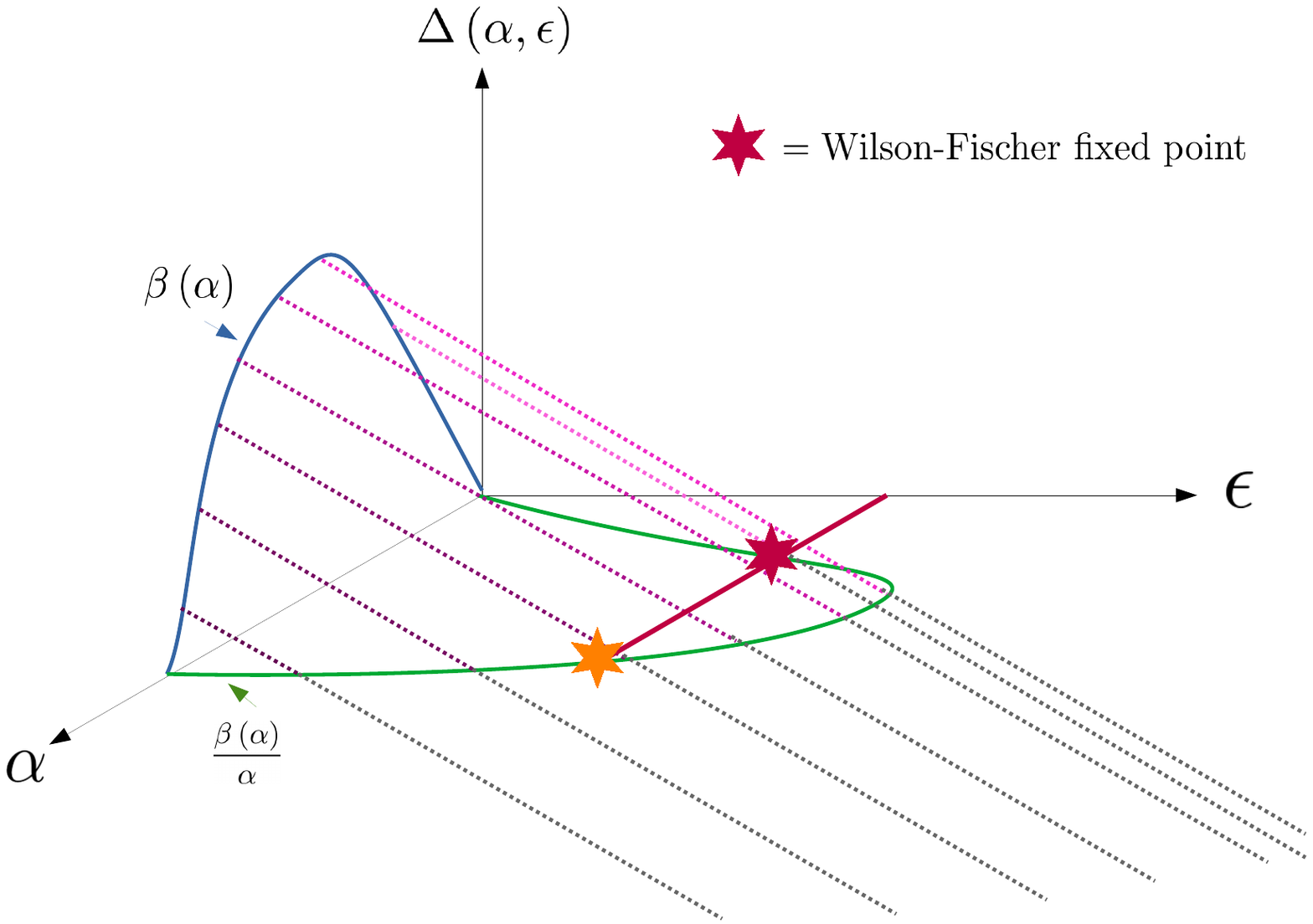}
    \caption{(Left) Sketch of our target beta function. (Right) 3D plot of $\Delta$. The pink lines follow the 3D surface of $\Delta$ for a fixed $\alpha$. The surface intersects the horizontal plane $\Delta=0$ along the green curve. For a fixed deformation $\epsilon$ there are two possible FPs; the red star corresponds to the Wilson-Fisher one while a second FP (orange star) is connected to the interacting FP in the $\epsilon\to 0$ limit.}
    \label{UV-IR model}
\end{figure}

We focus on a $\beta$-function featuring an interacting FP schematically shown in Fig.\ref{UV-IR model}a.

One can  draw the corresponding surface $\Delta(\alpha,\epsilon)$, plotted in Fig.\ref{UV-IR model}b, which for fixed $\alpha$ gives the corresponding line in the $\epsilon$ direction.

The intersection curve between the obtained surface and the $\Delta=0$ plane (plotted in green in Fig.\ref{UV-IR model}b) corresponds to all the possible FPs of the deformed theory in $d-\epsilon$ dimensions. This curve is defined via, the equation $\epsilon= \frac{\beta\left( \alpha\right)}{\alpha}=Y(\alpha)$.

\begin{figure}[h!]
    \centering
    \includegraphics[width=.49\textwidth]{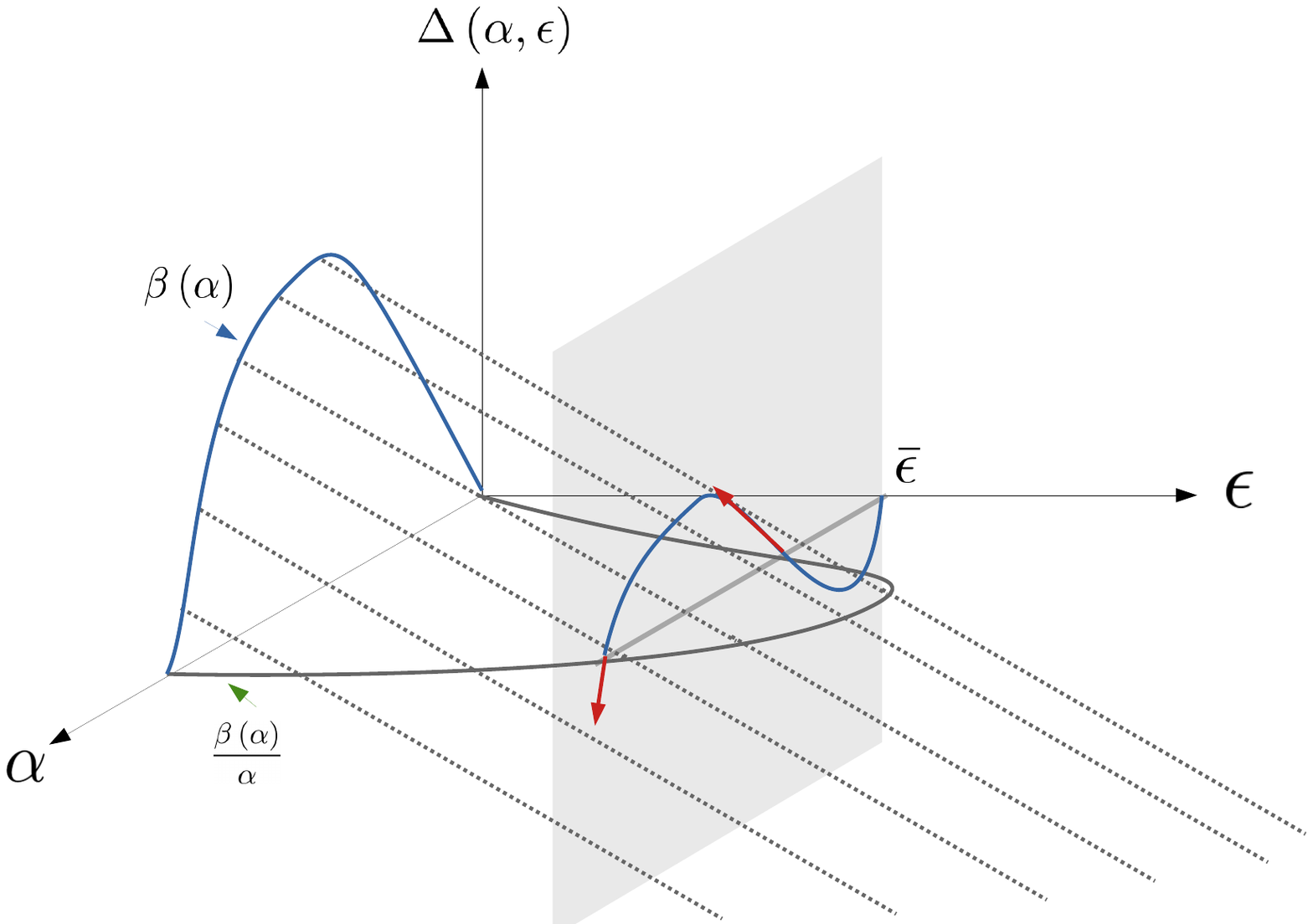}
    \includegraphics[width=.49\textwidth]{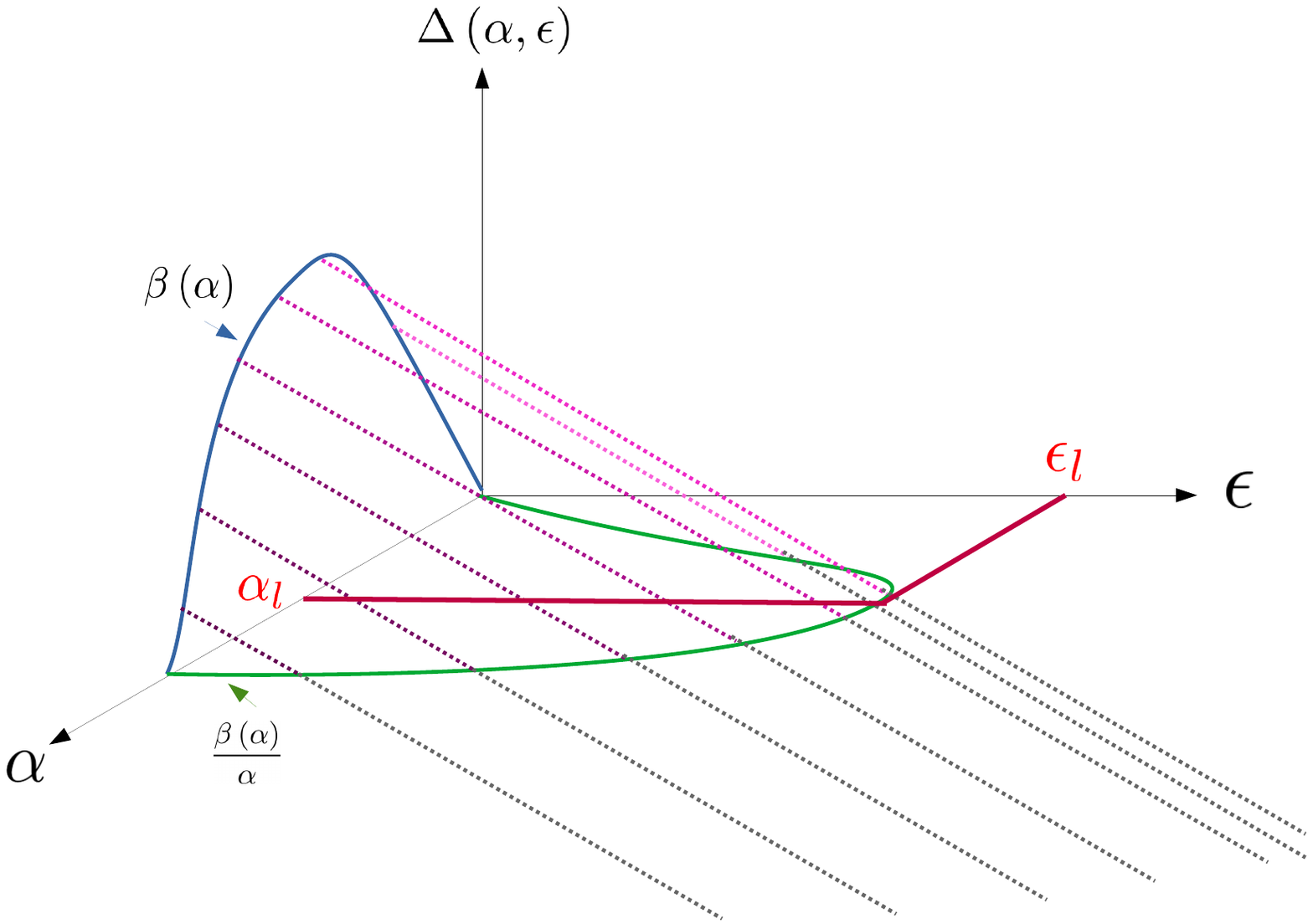}
    \caption{(Left) For a fixed deformation ($\bar{\epsilon}$ plane in grey), the tangent at the two fixed points are drawn in red. Manifestly they have to be of opposite signs. (Right) The merger point ($\alpha_l,\epsilon_l$).}
    \label{UV}
\end{figure}

We observe that for a given $\epsilon$ there are two possible FPs (highlighted by the red and orange stars in Fig.\ref{UV-IR model}b). The one close to the origin corresponds to the Wilson-Fisher FP (red star in Fig.\ref{UV-IR model}b), which is smoothly connected to the Gaussian FP in the $\epsilon \rightarrow 0$ limit. Moreover, as depicted in figure \ref{UV}a the two FPs must have their corresponding tangents with opposite signs leading to multi-valued $\omega$. Consequently at the merger $(\alpha_l, \epsilon_{l})$ the tangent must be zero i.e. $\omega\left(\epsilon_l\right)=0$ (see Fig.\ref{UV}b).
This merger point $(\alpha_l, \epsilon_{l})$ corresponds to the maximum of the curve $Y$, reached at $\alpha_l$ i.e. $Y\left(\alpha_l\right)=\epsilon_l$ and beyond this point one cannot define a Wilson-Fisher FP. In addition, the function $\omega$ should be \emph{non-Lipschitz} at this point otherwise (as we have shown in Fig.\ref{phasespace}) the solution of the SCR would behave asymptotically as a straight line for $\alpha\to \infty$ which will not match our target $\beta$-function.  

The zero of $\omega(\epsilon_l)=0$ is scheme invariant and we may confirm this by performing a scheme transformation $\alpha=\mathcal{G}\left(\Tilde{\alpha}\right)$ and compute the $\Tilde{Y}$-function in the new scheme. The details of the calculation can be found in appendix \ref{SchemeApp} where we show that $\Tilde{Y}\left(\Tilde{\alpha}\right)= Y\left(\alpha\right)$
and therefore for any $\mathcal{G}$, the maximum of $\Tilde{Y}$ will be identical to the maximum of $Y$. 

We conclude that the presence of an interacting FP in the undeformed theory requires another critical exponent which is smoothly connected to this FP in the $\epsilon \rightarrow 0$ limit. This second critical exponent might be unknown in practice but nevertheless necessary condition for the existence of the \emph{non-Lipschitz} point $\omega\left(\epsilon_l\right)=0$ can, in principle, be checked for the Wilson-Fisher branch. Also, non-Lipschitz behavior at the scheme-invariant merger would preclude the reconstruction of the $\beta$-function from SCR along the entire RG flow as we will see in explicit model in Sec.\ref{LS model}.

 \subsection{Poles} \label{polesection}
\begin{figure}[h!]
    \centering
    \includegraphics[trim={1cm 0cm 0 0},width=0.4\linewidth]{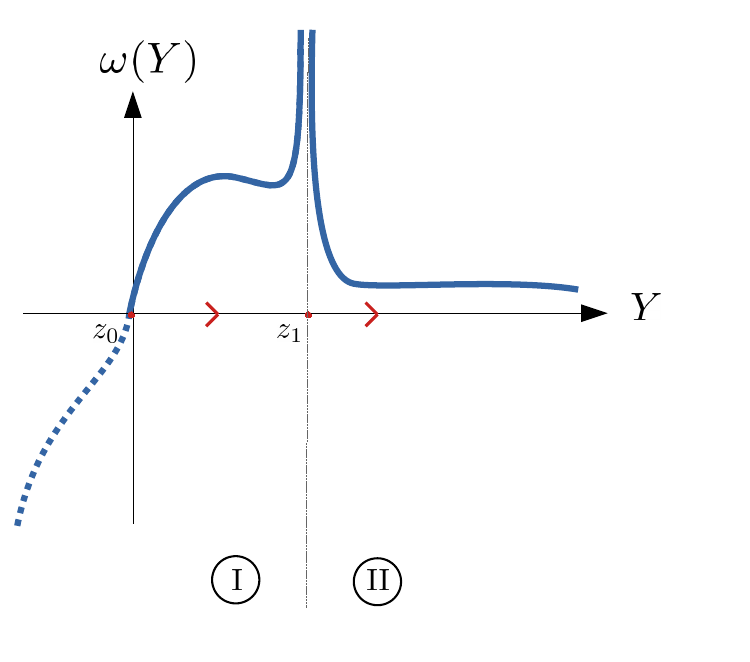} \hspace{1cm}
    \includegraphics[width=0.5 \linewidth]{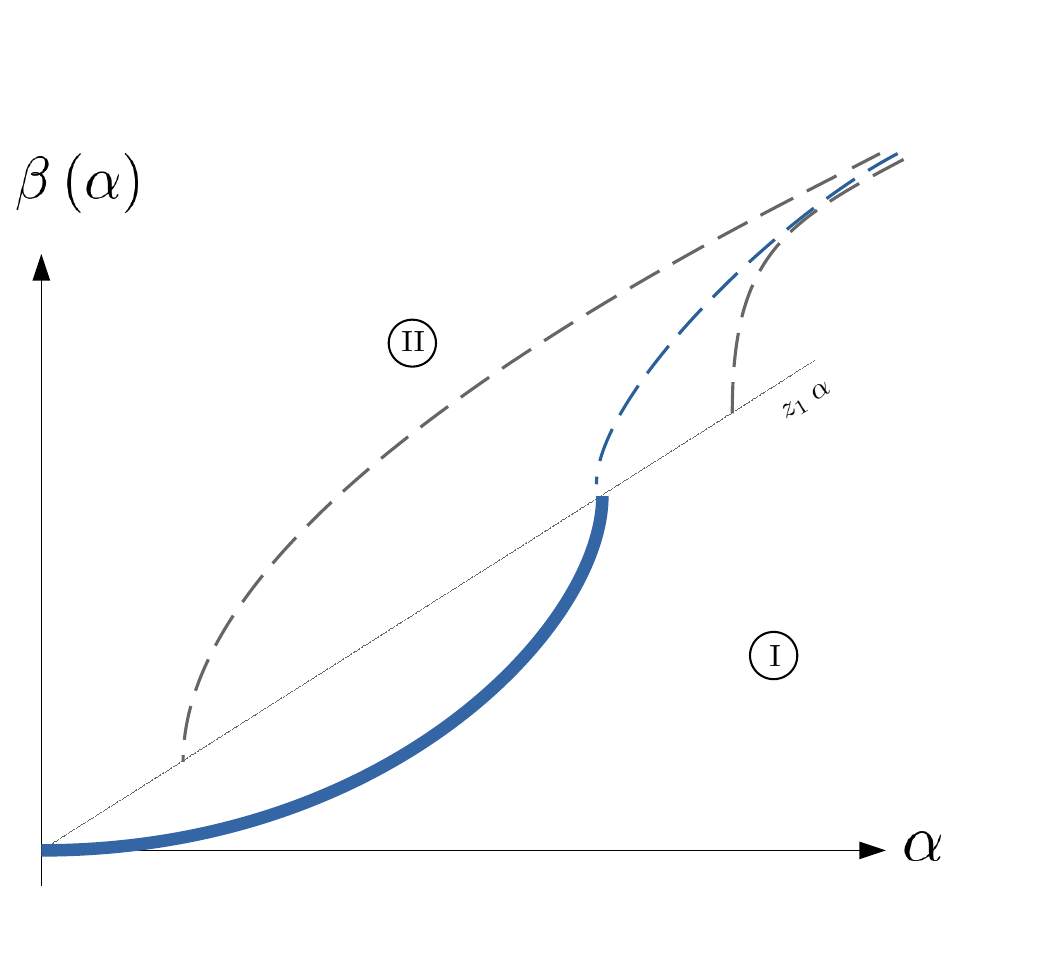}
    \vspace{0.5cm}
    \caption{(Left) Illustration of $\omega$ for the SCR with a pole localized at  $z_1$, delimiting the regions \MakeUppercase{\romannumeral 1} and  \MakeUppercase{\romannumeral 2}.  (Right) We draw the solutions taken with initial conditions in each region. We show several solutions (dashed) depending on the initial condition inside the region \MakeUppercase{\romannumeral 2}. The blue dashed solution is chosen to be continuously connected to the solution of the previous region.}
    \label{phasespacepolescenario1}
\end{figure}

\begin{figure}[h!]
    \centering
    \includegraphics[trim={1cm 0cm 0 0},width=0.4\linewidth]{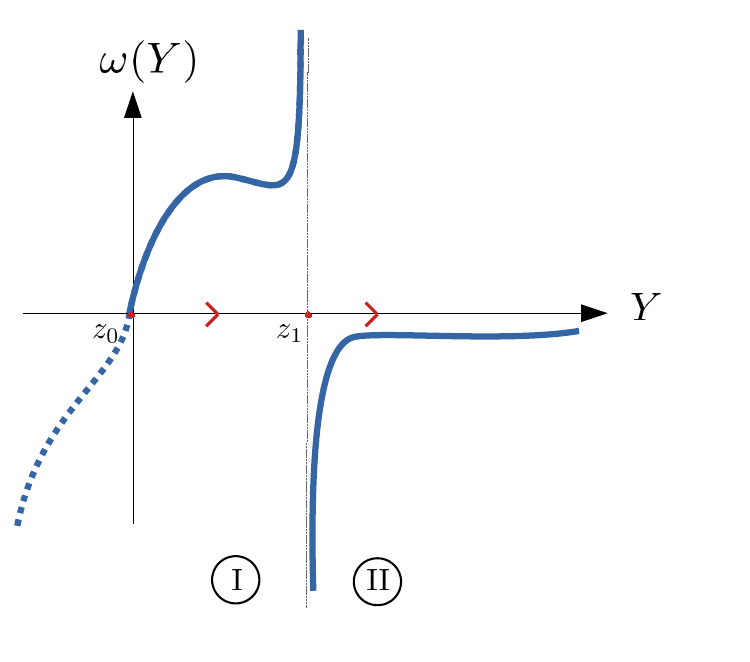} \hspace{1cm}
    \includegraphics[width=0.5 \linewidth]{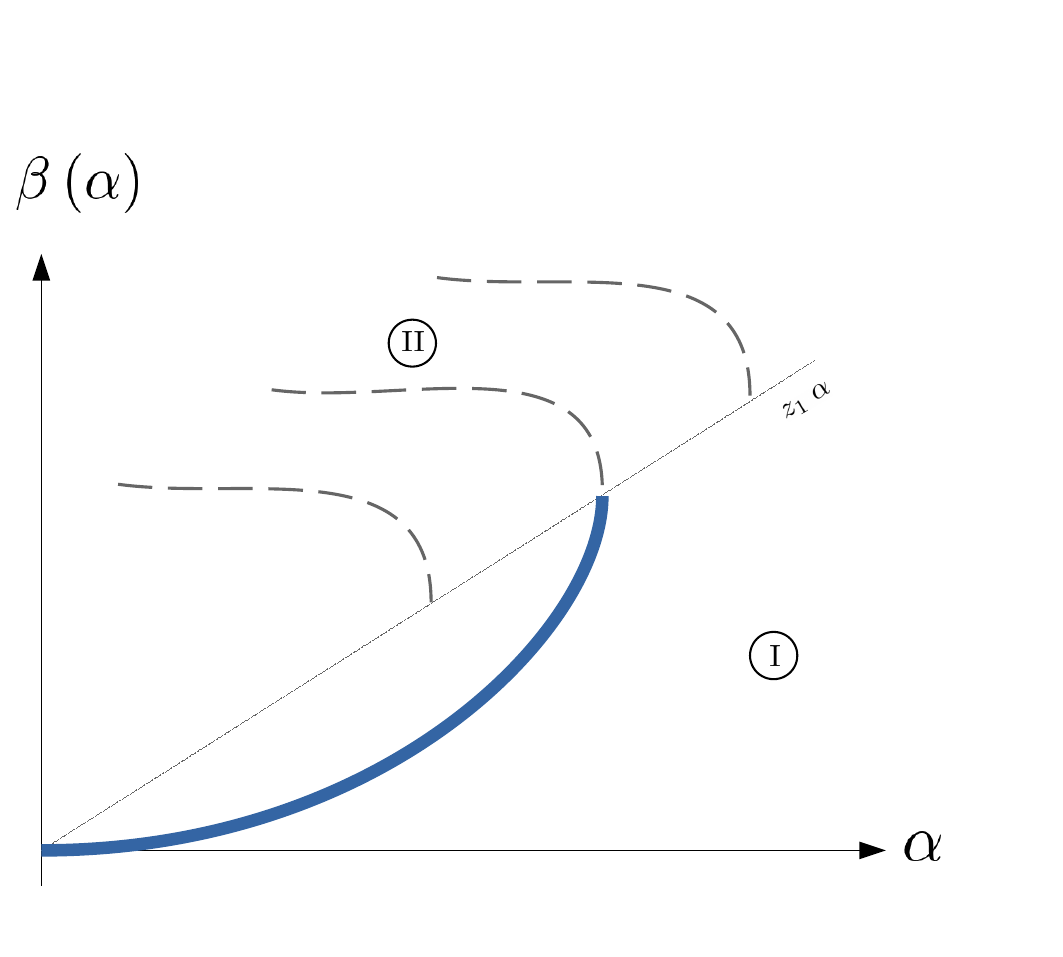}
    \vspace{0.5cm}
    \caption{(Left) Illustration of $\omega$ for the SCR with a pole localized at  $z_1$, delimiting the regions \MakeUppercase{\romannumeral 1} and  \MakeUppercase{\romannumeral 2} with opposite sign.  (Right) We draw the solutions taken with initial conditions in each region. We show several solutions (dashed) depending on the initial condition inside the region \MakeUppercase{\romannumeral 2}. }
    \label{phasespacepolescenario2}
\end{figure}

 The possible occurrence of a pole in $\omega$ needs to be treated with care. If the  sign is the same on both sides of the pole the resulting beta function can be analytically continued and remains singled valued in $\alpha$. However, it there a sign change across the pole the beta function, if continued, becomes double-valued and therefore leads to  non-physical behavior. The same sign behavior is illustrated in Fig.\ref{phasespacepolescenario1} while the sign-change is shown in  \ref{phasespacepolescenario2}.  We will see, in the \ref{largeNexamples} section, how poles emerge at large $N_f$  in gauge-fermion theories as well as at large $N$ in $O(N)$ models  . 

\section{Applications} \label{applications}

We now illustrate the general features of the SCR we discussed in the previous section in explicit examples. We start with the 4d gauge-fermion theories at the one and two-loop level where we emphasize the role of the initial conditions for the SCR, the zeros of $\omega$ and its non-Lipschitz behavior. Then we will turn to large $N$ models to illustrate the impact of the presence of poles in $\omega$. 
\subsection{Gauge-fermion Theories }

\subsubsection{1-loop QCD and QED}

Consider the one-loop \(\beta\)-function of a four-dimensional gauge-fermion theory with coupling \(\alpha\):
\[
\beta_{1-loop}(\alpha) \;=\; \beta_0\,\alpha^2,
\]
where \(\beta_0<0\) corresponds to asymptotically free (QCD-like) behavior, while \(\beta_0>0\) corresponds to QED-like theories. Inserting $\beta_{1-loop}$ into Eq.\ref{Beta} and Eq.\ref{omega}, the critical exponent is:
\begin{equation}
    \omega\left(\epsilon\right)=  \epsilon \ .
    \label{omegacritical}
\end{equation}

\begin{figure}[h!]
    \centering
    \includegraphics[trim={3cm 0cm 0 0},width=1.1\linewidth]{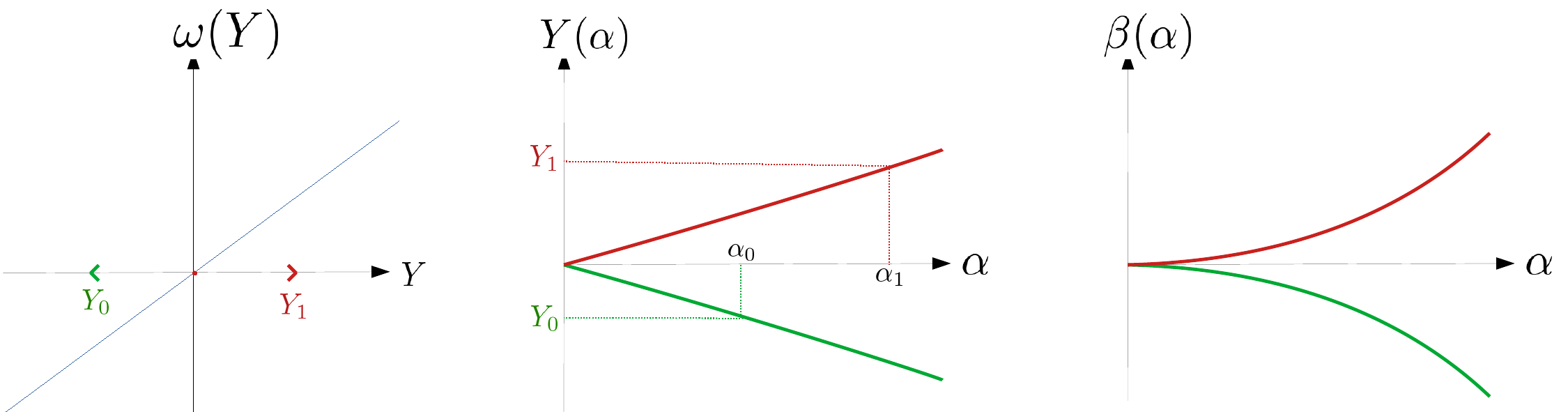}
    \caption{(Left) Plot of $\omega$ at 1-loop; the repulsive zero at the origin separates two different solutions. (Middle) Plot of the different solutions $Y$ of the SCR. (Right) Corresponding 1-loop $\beta$-function.}
    \label{QCD1loop}
\end{figure}

In the left panel of Fig.\ref{QCD1loop} we plot $\omega$ to highlight the repulsive zero at the origin $\omega\left(0 \right)=0$. The SCR then simplifies to:
\begin{equation}
    Y'\left(\alpha\right)=\frac{ Y\left(\alpha\right)} {\alpha} \ ,
\end{equation}
which can be solved analytically:
\begin{equation}
    Y(\alpha)    = Y_i \; \alpha /\alpha_i  \qquad \Longrightarrow  \qquad \beta(\alpha)= Y_i \;  \alpha^2/\alpha_i
\end{equation}
given an initial condition $Y(\alpha_i)=Y_i$ labeled by $i$. 

\noindent
A negative value for the initial condition, $Y_0 < 0$ in Fig.\ref{QCD1loop}, results in the asymptotically free QCD-like dynamics, while a positive initial condition, $Y_1 > 0$ in Fig.\ref{QCD1loop}, results in the QED one, demonstrating how initial conditions map into the first coefficient of the gauge-fermion beta-function coefficient.

\subsubsection{2-loops: Banks-Zaks fixed-point} \label{LS model}

To  highlight the impact of non-Lipschitz behaviour of the critical exponent, we consider a $SU(N_c)$ gauge theory with  $N_f$ Dirac fermions transforming into the fundamental representation. As customary, we will define the t'Hooft coupling $\kappa= \alpha N_c$ and take the Veneziano-Witten limit with $N_f$ and $N_c\rightarrow \infty$ while keeping their ratio fixed. We parametrize the distance from the asymptotic freedom boundary at $N_f/N_c=11/2$ via:

\begin{equation}
 \delta=  \frac{11}{2} -\frac{N_f}{N_c}   \ .
\end{equation}

The two-loops $\beta$-function reads:
\begin{equation}
    \frac{\mathrm{d} \kappa}{ \mathrm{d} \ln{\mu}} = -\frac{4}{3} \delta \kappa^2 +25 \kappa^3 + \mathcal{O}\left( \delta \kappa^3, \kappa^4  \right) 
    \label{LitimSannino}
\end{equation}
and since $\delta$ can be chosen arbitrary small, $0<\delta \ll 1$, the truncated $\beta$-function is under perturbative control. Eq.\ref{LitimSannino} reveals the existence of an IR Banks-Zaks FP \cite{Banks:1981nn} at $\kappa^*= \frac{4}{75} \delta$, which is also under perturbative control since $\delta \ll 1$.  
For the numerical results of this section we will use $\delta=0.1$.

\begin{figure}[h!]
    \centering
    \includegraphics[width=0.47\linewidth]{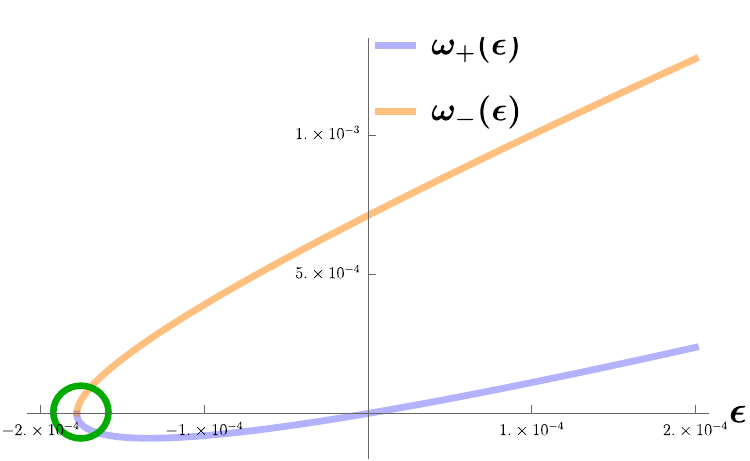}
    \hspace{1cm}\includegraphics[width=0.4\linewidth]{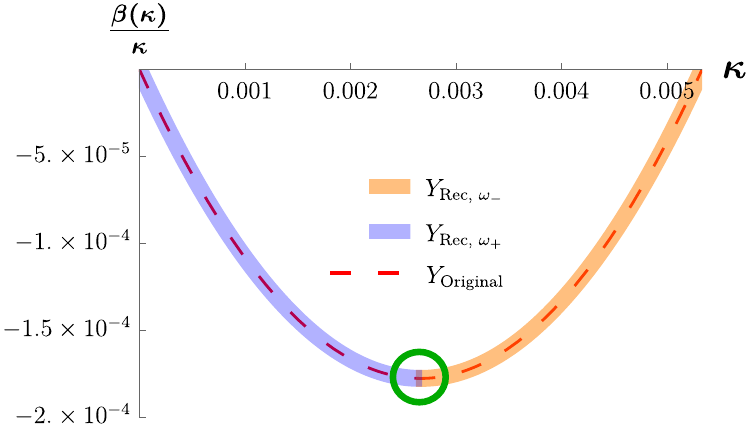}
    \caption{(Left) We show the two critical exponents corresponding to the two FPs of $\beta$, ($\delta=0.1$). In the $\epsilon\to 0$ limit, the blue solution is connected to the Gaussian FP while the orange one is connected to the Banks-Zaks FP. (Right) For both $\omega_\pm$, we solved numerically the SCR. The reconstructed solutions is plotted in terms of $Y$.}
    \label{recbeta}
\end{figure}

To extract the critical exponent we first find a zero of $\Delta\left(\kappa , \epsilon \right)= -\kappa \epsilon  -\frac{4}{3} \delta \kappa^2 +25 \kappa^3 $ and then insert it in Eq.\ref{omega}  to find the corresponding slope as a function of $\epsilon$. To make sure that we work in a well defined mathematical limit we take $\epsilon$ to scale like $\delta^2$. This guarantees the correct expansion in $\delta$ for the $d-\epsilon$ related theory. The two roots of $\Delta\left(\kappa , \epsilon \right)=0$ are $\kappa_\pm=  \frac{-1}{50} \left(-\frac{4 \delta }{3} \pm \sqrt{\frac{16 \delta ^2}{9}+100 \epsilon }  \right)$. The corresponding critical exponents, $\omega_+$ and $\omega_-$ are plotted in the left panel of Fig.\ref{recbeta} for $\delta = 0.1$ and their explicit form as function of $\delta$ and $\epsilon$ (assumed to be of order $\delta^2$) is given by:
\begin{equation}
    \omega_\pm\left(\epsilon\right)=2 \epsilon \pm \frac{-2}{75} \delta  \sqrt{\frac{16 \delta ^2}{9}+100 \epsilon }+\frac{8 \delta ^2}{225}  \ .
\end{equation}

Finally we solve SCR numerically for $\omega_+$ and $\omega_-$ separately. We plot the result for the reconstructed 
$Y_{\pm}$-functions by the corresponding blue and orange color in the right panel of Fig.\ref{recbeta}  together with the original input from Eq.\ref{LitimSannino} shown by red-dashed lines. The solution for $\omega_+$ matches the part smoothly connected to the Gaussian FP while the one or $\omega_-$ matches the part connected to the interacting IR fixed point and gluing them we reproduce the original $\beta$-function. This means that one alone is insufficient to reconstruct the full flow. 
The non Lipschitz behaviour of  $\omega$ is explicit in Fig.\ref{recbeta} occurring when $\omega_\pm$ vanish which is highlighted by the green circle.

\subsection{Large \texorpdfstring{$N$}{N}} \label{largeNexamples}

We now move to examples featuring a large number of degrees of freedom $N$ in which the 
  expansion can be organized in powers of $1/N$.

\subsubsection{  Gauge-fermion theory at the $1/N_f$ order} \label{1Ncontr}

Consider the 4d $SU(N_c)$ gauge theories at large number $N_f$ of fermionic matter fields. The $1/N_f$ expansion for the four dimensional $\beta$-function and the Wilson-Fisher $d-\epsilon $ critical exponent read \cite{Antipin:2018zdg}:
\begin{align}
\beta \left( x \right) &= \frac{2 x^2 }{3} \left[ 1 + \frac{F_1\left(x\right)}{N_f} + \frac{F_2\left(x\right)}{{N_f}^2} + \dots \right] \label{beta1N}\\
\omega\left(\epsilon\right)&=\epsilon+\sum_{n=1}^{\infty} \frac{\omega^{\left(n\right)}\left(\epsilon\right)}{{N_f}^n} \ .
\end{align}
Here the functions $F_n$ resum all perturbative orders in the 't Hooft coupling $x \equiv \alpha N_f T_R$ with $T_R=1/2$ for the fundamental representation. For our purposes it is sufficient to consider the latter case. 

\begin{figure}[h!]
    \centering
    \includegraphics[width=1\linewidth]{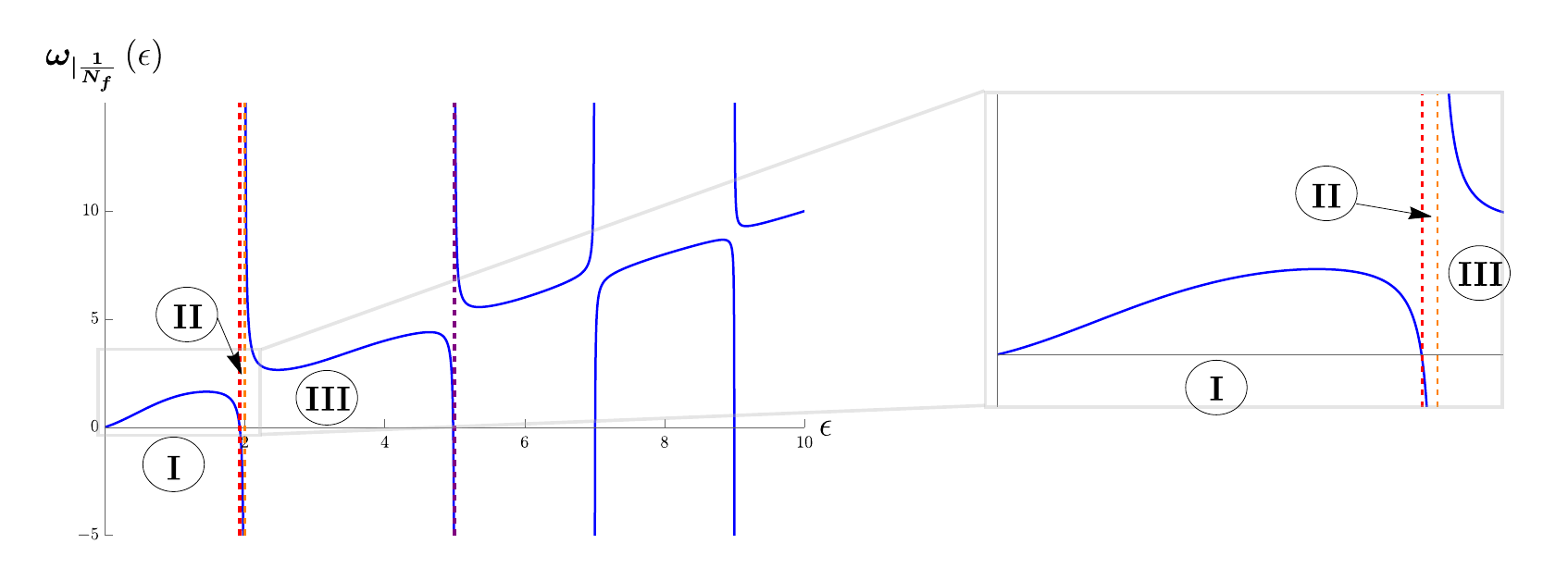}
    \includegraphics[width=1\linewidth]{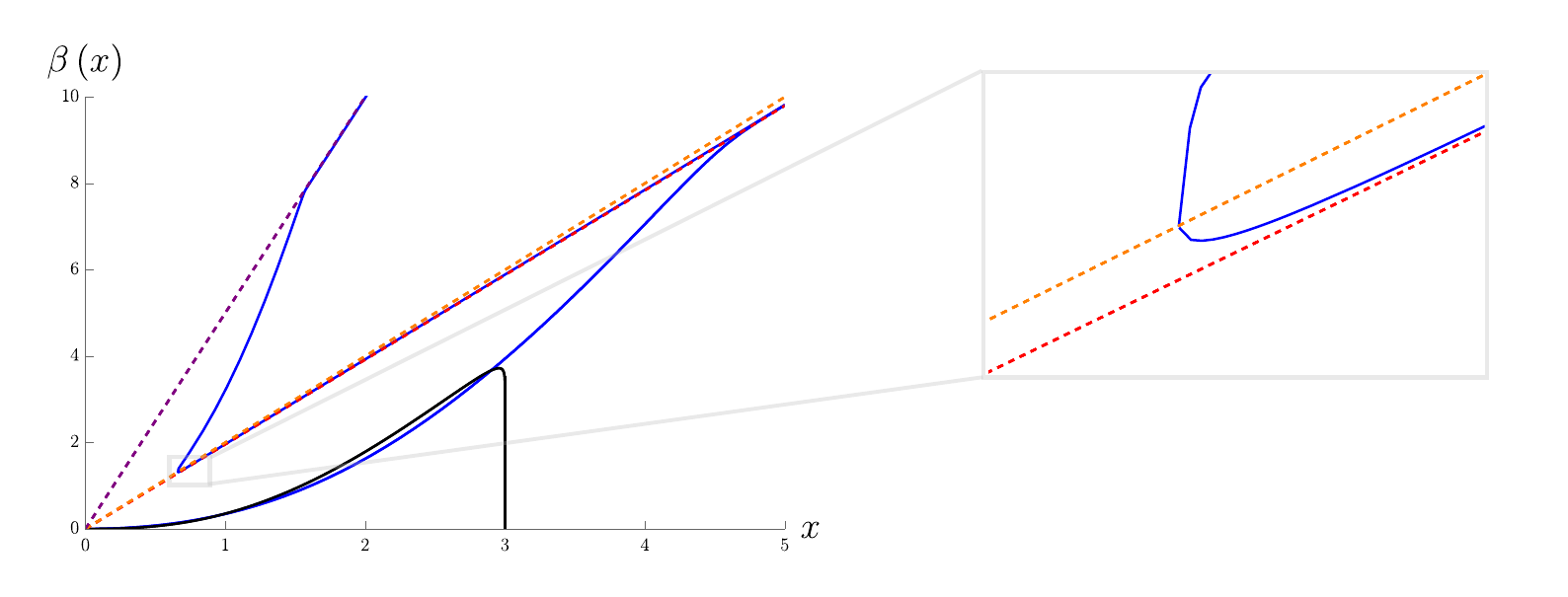}
    \caption{(Top) The critical exponent truncated at first order plotted on a wider interval for $N_f=20$ in a $SU(10)$ gauge theory. The dashed red and orange lines separate the first regions to account while solving the SCR. (Bottom) The SCR solutions for each region (blue) as well as the original $\beta_{|\frac{1}{N_f}}\left(x\right)=\frac{2 x^2 }{3} \left[ 1 + \frac{F_1\left(x\right)}{N_f}  \right]$ (black). }
    \label{blasi}
\end{figure}

The leading order function $F_1$ displays a negative pole  at $x=3$ implying that the $\beta$-function  restricted to this order ( $\beta_{|\frac{1}{N_f}}$ ), exhibits a UV FP \cite{Antipin:2017ebo,Holdom:2010qs,Palanques:1983ogz} as shown in black in the bottom-left panel of Fig.\ref{blasi}. We use as benchmark the $SU(10)$ gauge theory with $N_f=20$ Dirac fermions.  Beyond the $1/N_f$ limit it is not clear whether the fixed point survives. Nevertheless, preliminary negative lattice results exist for the $SU(2)$ case with 24 and 48 Dirac fermions \cite{Leino:2019qwk}.

 To gain some understanding on  this issue we will consider the impact on the four dimensional $\beta$-function by solving the SCR with $\omega^{\left(1\right)}$ as an input. This is interesting, as noticed in \cite{Alanne:2019vuk,Alanne:2019meg},
 because a fixed-order truncation in the critical exponent is not equivalent to the same-order
truncation in the $\beta$-function. In fact,   $\omega^{\left(1\right)}$ feeds partially to all the higher order functions $F_n$ as it is illustrated from the relations below, first for $F_1$  
\begin{equation}
    F_{1} \left(  x \right) =  \frac{3}{2}\int_0^x \frac{dt}{t^2}\omega^{\left(1\right)}( 2t/3 ) 
    \label{LO1N}
\end{equation}
and then for the higher order functions $F_n$ \cite{Alanne:2019vuk}:

\begin{equation}
    F_{n>1}\left(x\right) \supset \;\;\; \sim \int_0^x dt \; t^{n-3} \; \; \left[ \; F_1\left( t \right) \; \right]^{n-1} \;\;\frac{d^{n-1}}{dt^{n-1}} \left[ \omega^{(1)}(2t/3) \right] \ .
\end{equation}

Clearly, the partial contribution of $\omega^{\left(1\right)}$ to $F_n$ for $n \geq 2$ is not enough to recover them completely, as for such purpose one would need the other $\omega^{\left(k\right)}$ for $k \leq n$, yet, it is natural to sum all the $\omega^{\left(1\right)}$ contributions to the $\beta-$function to see their effect. 

In the top panel of Fig.\ref{blasi} we plotted $\omega_{|\frac{1}{N_f}}\left(\epsilon\right)=\epsilon+ \frac{\omega^{\left(1\right)}\left(\epsilon\right)}{N_f}$ and highlighted its poles. We notice that, except around the poles, it is well approximated by the $\epsilon$ term reconstructing the 1-loop term $\frac23 x^2$ of the beta-function in Eq.\ref{beta1N}    (compare also with Eq.\ref{omegacritical}). We also zoomed into the region around the first pole. Here region \MakeUppercase{\romannumeral 1} is defined from the origin to the first zero of $\omega^{\left(1\right)}$, region \MakeUppercase{\romannumeral 2} extends from this zero to the pole and region \MakeUppercase{\romannumeral 3} from the first pole to the second zero.  In the bottom panel of Fig.\ref{blasi} highlighted in blue we plot the corresponding solutions of the SCR for these three regions. We observe that due to existence of a zero at a finite value of the coupling for $\omega^{\left(1\right)}$ in the region \MakeUppercase{\romannumeral 1} the solution asymptotes to a straight line for $x\to +\infty$ while in regions \MakeUppercase{\romannumeral 2} and \MakeUppercase{\romannumeral 3} it is double-valued due to the different sign of $\omega^{\left(1\right)}$ across the singularity. This is best seen from the zoom into the narrow region \MakeUppercase{\romannumeral 2}. Because of the multi-valued nature of the resulting beta function no   conclusions can be drawn about the fate of the $1/N_f$ UV FP.

\subsubsection{O(\texorpdfstring{$N$}{N}) model at order $1/N^2$ \label{sec:O(N)}}
 As another relevant example we investigate the O($N$) mode to the next-to-leading order corrections in $1/N$ \cite{Gracey:1993kb}\cite{Gracey:1996ub}.

\begin{figure}[!ht]
    \centering
    \includegraphics[width=0.75
    \linewidth]{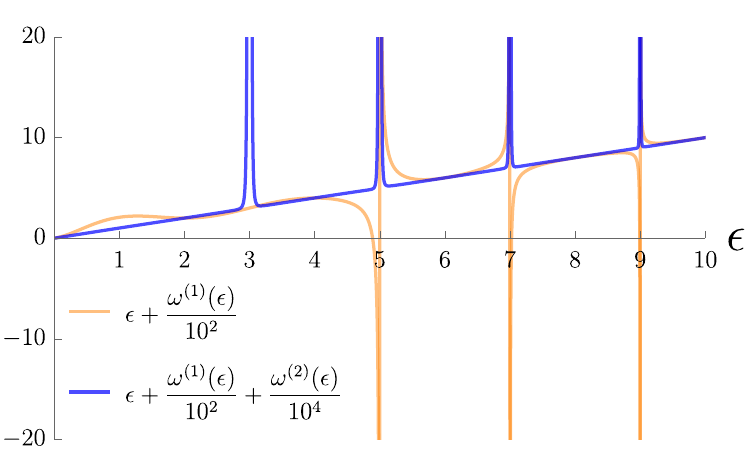}
    \caption{The $O(N)$ critical exponent truncated at first order in orange and second order in blue.}
    \label{o(n)}
\end{figure}

In Fig.\ref{o(n)} we plot, in orange, the $1/N$ contribution to the critical exponent for $N=100$. This features zeros and poles similarly to the gauge theory example above.  However, the \,\(\mathcal{O}(1/N^2)\) corrections feature stronger singularities so their addition allows to change 
the sign around the $1/N$ poles (shown by the blue curve in Fig.\ref{o(n)}) permitting to obtain a single-valued solution for the four-dimensional beta function.  

\begin{figure}[h!]
    \centering
    \includegraphics[width=0.475\linewidth]{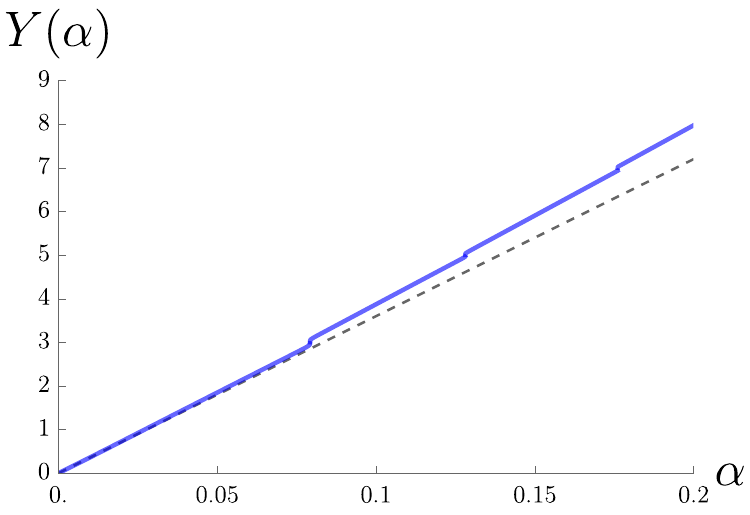 }
    \includegraphics[width=0.475\linewidth]{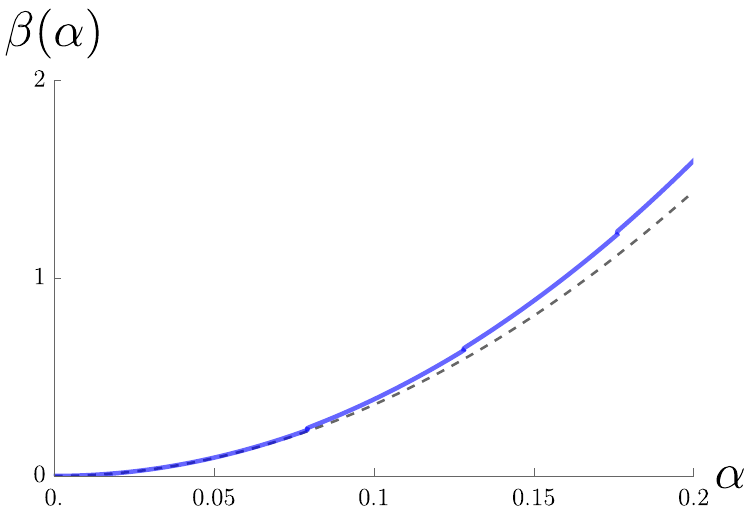}
    \caption{Solutions of the SCR for the truncated critical exponent at second order (left $Y$-function and right $\beta$-function). The apparent "kicks" correspond to the solutions getting closer to the poles and hence delimit the different intervals over which the SCR is numerically solved. The black dashed line is the 1-loop $Y$($\beta$)-function.}
    \label{o(n)rec}
\end{figure}

In Fig.\ref{o(n)rec} we show the reconstructed solutions where each "kick" corresponds to the solutions getting closer to a pole delimiting the region where the SCR is numerically solved. Within each interval, we had to adjust the initial condition to glue the solution to the previous region. We observe that our result is close to the 1-loop $\beta$-function shown in dashed line since the truncated critical exponent at $1/N^2$ order is well approximated by the $\epsilon$ term  (compare with Eq.\ref{omegacritical}) except around the poles just. This feature is similar to the gauge-fermion example above.

Based on this result, we may speculate on the effect of $1/N^2$ corrections in the gauge theory above. These corrections may similarly change the sign of the singularities in which case the solutions of the SCR away from the poles will be close to the 1-loop beta function as in the $O(N)$ example while if they do not change  sign the reconstructed beta function will be multi-valued, and hence ill-defined.

\section{Conclusion} \label{conclusions}

We investigated the critical behavior of generic QFTs featuring a single marginal coupling in $d-\epsilon$ dimensions to extract information about its $d$ dimensional counterpart. To achieve this goal we first derived and then analyzed the implications of the self-consistent relation between the $d-\epsilon$ scaling exponent and the $d$ dimensional beta function. The relation allowed us show that to describe an interacting FP in the $d$ dimensional theory at least two critical exponents must exist as function of $\epsilon$. These eventually collide for a given $\epsilon$ in a non-Lipschitz way. We illustrated this feature for the Banks-Zaks infrared FP in 4$d$ QCD. However, for the large number of flavours limit in QCD the absence of a multivalued critical exponent for the $d-\epsilon$ sister theory prevents from concluding on the absence of a non-trivial UV fixed point. This result shows that the statement made in \cite{Alanne:2019vuk} is invalid. 
Moreover, we demonstrated that the potential existence of poles in the critical exponents as function of $\epsilon$ in the $d-\epsilon$ theory prohibits the reconstruction of the full RG flow, when the critical exponent signs on both sides of the pole are opposite. In fact, if the signs agree one can reconstruct the flow by gluing together different regions of the RG flow. Interestingly, we observed that, at the order $1/N$, the time-honored four-dimensional $O(N)$ model features poles with switching signs for the critical exponents of the sister theory in $d-\epsilon$ dimensions. However, the full flow for the theory can be reconstructed when taking into account the $1/N^2$ corrections. We expect a similar behavior to occur also for gauge-fermion theories at large number of flavors.  It would be interesting to generalize our results to multi-coupling theories.

\section*{Acknowledgements}

\paragraph{Funding information}  SV is supported by the Fonds de la Recherche Scientifique de Belgique (FNRS)  under the IISN convention 4.4517.08 . AP is a boursier of Université Libre de Bruxelles and supported by the EoS FNRS grant "Beyond Symplectic Geometry". The work of F.S. is partially supported by the Carlsberg Foundation, grant CF22-0922.

\begin{appendix}

\section{Scheme Transformation \label{SchemeApp}}

After the scheme transformation $\alpha=\mathcal{G}\left(\Tilde{\alpha}\right)$, our $\Delta$-function from Eq.\ref{Beta} is modified to $\Tilde{\Delta}$ according to:

\begin{equation}
    \Tilde{\Delta}\left( \Tilde{\alpha}, \epsilon \right) = \frac{\mathrm{d} \; \Tilde{\alpha}}{\mathrm{d} \ln{\mu}} = \frac{\mathrm{d} \; \Tilde{\alpha}}{\mathrm{d} \alpha } \frac{\mathrm{d} \; {\alpha}}{\mathrm{d} \ln{\mu}}=\frac{\Delta\left( \mathcal{G}\left( \Tilde{\alpha}\right),\epsilon\right)}{\mathcal{G'}\left( \Tilde{\alpha}\right)}=-\frac{ \mathcal{G}\left( \Tilde{\alpha}\right) }{\mathcal{G'}\left( \Tilde{\alpha}\right)}\epsilon + \frac{ \beta\left( \mathcal{G}\left( \Tilde{\alpha}\right) \right)}{\mathcal{G'}\left( \Tilde{\alpha}\right)} \ .
\end{equation}

The $\epsilon \rightarrow 0$ limit gives us the  new $d-$dimenional beta function $\Tilde{\beta}$: 

\begin{equation}
    \Tilde{\beta}\left( \Tilde\alpha\right) =  \frac{ \beta\left( \mathcal{G}\left( \Tilde{\alpha}\right) \right)}{\mathcal{G'}\left( \Tilde{\alpha}\right)}
\end{equation}
so, as a consequence, the $\tilde{Y}$-curve which is given by the equation  $\Tilde{\Delta}\left(\Tilde{\alpha},\epsilon\right)=0$ in the $\left(\Tilde{\alpha},\epsilon\right)$ plane becomes:

\begin{equation}
  \Tilde{\Delta}\left(\Tilde{\alpha},\epsilon\right)=-\frac{ \mathcal{G}\left( \Tilde{\alpha}\right) }{\mathcal{G'}\left( \Tilde{\alpha}\right)}\epsilon + \frac{ \beta\left( \mathcal{G}\left( \Tilde{\alpha}\right) \right)}{\mathcal{G'}\left( \Tilde{\alpha}\right)}=0 \;\; \Rightarrow \;\; \epsilon=  \frac{ \beta\left( \mathcal{G}\left( \Tilde{\alpha}\right) \right)}{\mathcal{G'}\left( \Tilde{\alpha}\right)} \frac{ \mathcal{G'}\left( \Tilde{\alpha}\right) }{\mathcal{G}\left( \Tilde{\alpha}\right)} \;\; \Rightarrow \;\; \Tilde{Y}\left(\Tilde{\alpha}\right)= \frac{ \beta\left( \mathcal{G}\left( \Tilde{\alpha}\right) \right)}{\mathcal{G}\left( \Tilde{\alpha}\right)} \ .
\end{equation}

Subsequently, 

\begin{equation}
    \Tilde{Y}\left(\Tilde{\alpha}\right)= Y\left(\alpha\right)
\end{equation}
and therefore, the maximum of the $\tilde{Y}$-curve is indeed scheme invariant, and the value $\Tilde{\alpha}_l$ where the maximum is reached is scheme "covariant".

\end{appendix}



\bibliography{SciPost_Example_BiBTeX_File.bib}


\end{document}